\magnification 1200
\def\[{[\![}
\def\]{]\!]}
\def\Z{{\bf Z}}
\def\C{{\bf C}}

\def\l{\ldots}
\def\n{\noindent}
\def\h{\hat{h}}
\def\e{\hat{e}}
\def\f{\hat{f}}
\def\n{\noindent}
\def\t{\theta}
\def\s{\smallskip}
\def\b{\bigskip}
\def\a{\hat{a}}

\def\L{{\bar L}}
\def\q{{\bar q}}

\def\d{\delta}
\def\v{\varepsilon}
\def\ra{\rangle}
\def\la{\langle}
\def\bu{\bullet}
\def\o{\circ}

\b\b

\baselineskip=16pt

\font\twelvebf=cmbx12

\b\b

\centerline{\twelvebf Fock representations of the superalgebra 
sl(n+1$|$m), its quantum}
\centerline{\twelvebf analogue U$_{\bf q}$[sl(n+1$|$m)]
and related quantum statistics}

\leftskip 36pt

\vskip 40pt
\noindent
T.D. Palev$^{\bf a)}$, N.I. Stoilova
$^{\bf b)}$\footnote*{\rm Permanent address: Institute for Nuclear 
Research and Nuclear Energy, Boul. Tsarigradsko Chaussee 72,
1784 Sofia, Bulgaria, (e-mail: stoilova@inrne.bas.bg).} 
and J. Van der Jeugt$^{\bf c)}$\footnote{**}{\rm 
Research Associate of the Fund for Scientific Research - Flanders 
(Belgium);
(e-mail: Joris.VanderJeugt@rug.ac.be).}

\noindent 
$^{\bf a)}$Institute for Nuclear Research and Nuclear 
Energy, Boul. Tsarigradsko Chaussee 72,
1784 Sofia, Bulgaria, e-mail: tpalev@inrne.bas.bg

\n$^{\bf b)}$Abdus Salam International Centre for Theoretical Physics,
P.O. Box 586, 34100 Trieste, Italy

\n$^{\bf c)}$Department of Applied Mathematics and Computer
Science, University of Ghent

\n Krijgslaan 281-S9, B-9000 Gent, Belgium

\vskip 54pt

\n{\bf Abstract.} Fock space representations of the Lie
superalgebra $sl(n+1|m)$ and of its quantum analogue
$U_q[sl(n+1|m)]$ are written down. The results are based on a
description of these superalgebras via creation and annihilation
operators. The properties of the underlying statistics are
shortly discussed.

\b\b\b

\vfill \eject

\vskip 48pt

\leftskip 0pt

\n
{\bf 1. Introduction}

\bigskip\n
The quantization of the  simple  Lie algebras [1]
and the  basic Lie superalgebras  [2] is usually carried out in terms
of their Chevalley generators $e_i, f_i, h_i,  i=1,\ldots,n$
for an algebra of rank $n$. Recently it was realized
that the quantum algebras $U_q[osp(1|2n)]~  [3],~ U_q[so(2n+1)] ~
[4]$,~ $ U_q[osp(2r+1|2m)], r+m=n ~[5],~ {\rm
and~~also}~~U_q[sl(n+1)] $ ~[6] and $U_q[sl(n+1|m)]$ ~[7] admit 
a description in terms of an alternative set of generators
$a_i^\pm,\; H_i,\;\;i=1,\ldots,n,$ referred to as (deformed)
creation and annihilation operators (CAOs) or generators. This
certainly holds also for the corresponding nondeformed Lie
superalgebras.

The concept of creation and annihilation operators
of a simple Lie (super)algebra was introduced in [8].
Let ${\cal A}$ be such an algebra with a
supercommutator $\[\;,\;\]$. The root vectors
$a_1^{\xi},\l,a_n^{\xi}$ of  ${\cal A}$ are said to be creation
($\xi=+$) and annihilation ($\xi=-$) operators of  ${\cal A}$, if 
$$
{\cal A}=lin.env.\{a_i^{\xi},\;\[a_j^{\eta},a_k^{\varepsilon}\]\;|\;
i,j,k=1,\l,n;\; \xi, \eta, \varepsilon =\pm \}, \eqno(1)
$$
so that $a_1^+,\l,a_n^+~~~~ ({\rm resp.}~~ a_1^-,\l,a_n^-)$
are negative (resp. positive) root vectors of  ${\cal A}$.

The Fock representations of ${\cal A}$, defined in [8], are
constructed in a much similar way as those of Bose or Fermi
operators (or their generalizations, the parabosons and
parafermions [9]). In a more mathematical terminology the
Fock modules of  ${\cal A}$ are induced from trivial
one-dimensional modules of a subalgebra ${\cal B}$ of  ${\cal A}$,
namely
$$
{\cal B}=lin.env.\{a_i^-,\;\[a_j^{\eta},a_k^{\varepsilon}\]\;|\;
i,j,k=1,\l,n;\; \eta, \varepsilon =\pm \}. \eqno(2)
$$

The reason  for  introducing such (more physical) terminology is
based on the observation that the creation and the annihilation
operators of the orthosymplectic Lie superalgebra (LS)
$osp(2r+1|2m)$ have a direct physical significance:
$a_1^\pm,\ldots,a_m^\pm~ ({\rm resp.}~~
a_{m+1}^\pm,\ldots,a_n^\pm)$ are para-Bose (resp.  para-Fermi)
operators [10], operators which generalize  the statistics of
tensor (resp. spinor) fields in quantum field theory
(for $n=m=\infty$) [9]. Since $osp(2r+1|2m)$ is an algebra from the
class $B$ in the classification of Kac [11] one may call the
paraquantization a $B$-quantization.

It was argued in [12] that to each class $A$, $B$, $C$ and $D$ of
basic LSs  there corresponds a quantum statistics, so that their
CAOs can be interpreted as creation and annihilation operators of
(quasi)particles, excitations, in the corresponding Fock
space(s).  This assumption holds for the classes $A$~, $B$~,
$C$~, $D$ of simple Lie algebras [13]. It was studied in detail
for the Lie algebras $sl(n+1)$ ($A-$statistics) [14], for the LSs
$sl(1|m)$ ($A-$superstatistics) [8, 15] and recently for the
classical Lie superalgebra $q(n+1)$ [16].

In the present note we report shortly on the properties of 
quantum statistics, related to the Lie superalgebra $sl(n+1|m)$
and its quantum analogue. These statistics are particular kinds
of generalized quantum statistics. There are many publications on
the subject especially in the part related to quantum groups
(initiated in [17]). Other interesting approaches to
generalized statistics and their Fock space representations are
also available. We mention here the generalizations associated
with the spectral quantum Yang-Baxter equations [18], the statistics 
based on Lie supertriple systems [19] and on triple operator algebras
[20], or extended Haldane statistics [21]. In certain approaches, one 
starts with a deformation of the Fock space, and from here one 
deduces the properties of creation and annihilation operators and 
the related statistics, see e.g.~Ref.~[22], or Ref.~[23] for
orthobose and orthofermi statistics.

For a selfconsistency of the exposition we recall in Sect. 2 the
description of $sl(n+1|m)$ and $U_q[sl(n+1|m)]$ via (deformed)
creation and annihilation operators as given in [7]. The
transformations of the Fock modules under the action of the CAOs
are written down in Sect. 3. In the last section we discuss
shortly the underlying statistics of the creation and
annihilation operators.

Throughout the paper we use the notation:

\smallskip
LS, LS's - Lie superalgebra, Lie superalgebras;

CAOs - creation and annihilation operators;

lin.env. - linear envelope;

$\Z$ - all integers;

$\Z_+$ - all non-negative integers;

$\Z_2=\{\bar{0},\bar{1}\}$ - the ring of all integers modulo 2;

$\C$ - all complex numbers;

$[p;q]=\{p,p+1,p+2,\l,q-1,q\}$, for $p\le q\in \Z $;\hfill (3)

\smallskip
$
\t_i=\cases {{\bar 0}, & if $\; i=0,1,2, \ldots , n$,\cr 
               {\bar 1}, & if $\; i=n+1,n+2,\ldots ,n+m$,\cr }; \quad
\t_{ij}=\t_i+\t_j; \hfill (4)
$

\smallskip
$
[a,b]=ab-ba,\;\; \{a,b\}=ab+ba,
\;\;\[a,b\]=ab-(-1)^{\deg(a)\deg(b)}ba; \hfill (5)
$

$
[a,b]_x=ab-xba,\;\; \{a,b\}_x=ab+xba,
\;\;\[a,b\]_x=ab-(-1)^{\deg(a)\deg(b)}xba. \hfill (6)
$

\smallskip
\bigskip\n
{\bf 2. The Lie superalgebra sl(n+1$|$m) and its quantization 
U$_q$[sl(n+1$|$m)]}

\bigskip\n
The universal enveloping algebra  $U[gl(n+1|m)]$ of the general 
linear LS
$gl(n+1|m)$ is a $\Z_2-$graded associative unital algebra 
generated by $(n+m+1)^2\;$ $\Z_2-$graded indeterminates 
$\{e_{ij}|i,j\in [0;n+m]\}$,\quad  $\deg(e_{ij})=\t_{ij}\equiv 
\t_i+\t_j,$
subject to the relations
$$
\[e_{ij},e_{kl}\]=\d_{jk}e_{il}-(-
1)^{\t_{ij}\t_{kl}}\d_{il}e_{kj}\quad
i,j,k,l=1,\ldots , n+m. \eqno(7)
$$ 
The LS $gl(n+1|m)$ is a subalgebra of $U[gl(n+1|m)]$, considered
as a Lie superalgebra, with generators  $\{e_{ij}|i,j\in [0;n+m]\}$ 
and supercommutation relations (7). The LS $sl(n+1|m)$ is a
subalgebra of $gl(n+1|m)$:
$$sl(n+1|m)=lin.env.\{e_{ij}, (-1)^{\t_k}e_{kk}-(-1)^{\t_l}e_{ll}|
i\ne j;\;  i,j,k,l\in [0; n+m]\}. \eqno(8)$$

\n
The generators $e_{00},~e_{11},~\l,~e_{n+m,n+m}$
constitute a basis in the Cartan subalgebra of 
$gl(n+1|m)$. Denote by $\v_0,~\v_1,~\l,~\v_{n+m}$
the dual basis, $\v_i(e_{jj})=\d_{ij}$. The root vectors of both 
$gl(n+1|m)$ and $sl(n+1|m)$ are $e_{ij}, \; i\ne j,\;i,j\in [0;n+m]$.
The root corresponding to $e_{ij}$
is $\v_i-\v_j$. With respect to the natural order of the basis in the 
Cartan subalgebra $e_{ij}$ is a positive (resp. a negative) root 
vector if $i<j$ (resp. $i>j$).

This  description of $sl(n+1|m)$ is simple, but it is not
appropriate for quantum deformations. Another definition
is given in terms of the Chevalley generators
$$
\h_i=e_{i-1,i-1}-(-1)^{\t_{i-1,i}}e_{ii},\quad \e_i=e_{i-1,i}, \quad
\f_i=e_{i,i-1},\;  i\in[1;n+m] \eqno(9) $$

\n
and the $(n+m)\times (n+m)$ Cartan matrix $\{\alpha_{ij}\}$ with
entries
$$
\alpha_{ij}=(1+(-1)^{\t_{i-1,i}})\delta_{ij}-
(-1)^{\t_{i-1,i}}\delta_{i,j-1}-\delta_{i-1,j},\;\;
i,j\in [1;n+m].\eqno (10) 
$$

\n
$U[sl(n+1|m)]$ is an associative unital algebra of the
Chevalley generators, subject to the Cartan-Kac and the Serre
relations:
$$
\eqalignno{
& [\h_i,\h_j]=0,\quad
[\h_i,\e_j]=\alpha_{ij}\e_j,&\cr
& [\h_i,\f_j]=-\alpha_{ij}\f_j,\quad
\[\e_i,\f_j\]=\delta _{ij}\h_i, & (11) \cr 
}
$$
$$
\eqalignno{
& [\e_i, \e_j]=0, \quad  [\f_i, \f_j]=0,\quad
\hbox{if }\; |i-j|\neq 1;& (12a) \cr
& \e^2_{n+1}=0,\quad \f^2_{n+1}=0;  & (12b)\cr
& [\e_i, [\e_i, \e_{i+1}]]=0,\quad [\f_i, [\f_i, \f_{i+1}]]=0,
\quad i\ne n+1;& (12c) \cr
& [\e_{i+1}, [\e_{i+1}, \e_{i}]]=0,\quad [\f_{i+1}, [\f_{i+1}, 
\f_{i}]]=0,
\quad i\ne n;& (12d) \cr
& \{[\e_{n+1}, \e_n],[\e_{n+1}, \e_{n+2}]\}=0,\quad
 \{[\f_{n+1}, \f_n],[\f_{n+1}, \f_{n+2}]\}=0. & (12e) \cr
}
$$
The grading on $U[sl(n+1|m)]$ is induced from the requirement
that the only odd generators are $\e_{n+1}$ and $\f_{n+1}$.
The LS $sl(n+1|m)$ is a subalgebra of  $U[sl(n+1|m)]$, generated 
by the Chevalley generators in the sense of a Lie superalgebra.

Consider the following root vectors from $sl(n+1|m)$:
$$
\a_i^+=e_{i0}, \quad  \a_i^-=e_{0i}, \;\;i\in [1;n+m], \eqno(13a)
$$ 
or, equivalently 
$$
\eqalignno{
& \a_1^-=\e_1,\quad 
\a_i^-=[[[\l[[\e_1,\e_2],\e_3],\l],\e_{i-1}],\e_i]=[\a_{i-1}^-, e_i],
\quad i\in [2;n+m],   & \cr
&\a_1^+=\f_1, \quad 
\a_i^+=[\f_i,[\f_{i-1},[\l ,[\f_3,[\f_2,\f_1]]\l]]]=[f_i,\a_{i-1}^+].
\quad i\in [2;n+m].   & (13b)\cr
}
$$
The root of $\a_i^-$ (resp. of $\a_i^+$) is $\v_0-\v_i$ 
(resp. $\v_i-\v_0$). Therefore (with respect to the natural
order of the basis $\v_0,\v_1,\l,\v_{n+m}$) 
$\a_1^-,~\l,~\a_{n+m}^-$~~ are positive root  vectors,
and $\a_1^+,\l,~\a_{n+m}^+$~~ are negative root vectors.
Moreover
$$
sl(n+1|m)=lin.env.\{\a_i^{\xi},\;\[\a_j^{\eta},\a_k^{\varepsilon}\]\;|
i,j,k\in[1;n];\; \xi, \eta, \varepsilon =\pm \}. \eqno(14)
$$
Hence, the generators (13) are creation and annihilation operators of 
$sl(n+1|m)$.
These generators satisfy the following triple relations:
$$
\eqalignno{
& \[\a_i^\xi ,\a _j^\xi \]=0, \quad \xi=\pm, \quad i,j\in [1;n+m],
&  \cr
& \[ \[ \a_i^+ ,\a _j^- \], \a _k^-\]=-(-1)^{\t_{ij}\t_k}\delta_{ik}
\a _j^- -
(-1)^{\t_i}\delta_{ij}\a _k^-,
& (15)\cr
& \[ \[ \a_i^+ ,\a _j^- \], \a _k^+\]=\delta_{jk}\a _i^++
(-1)^{\t_i}\delta_{ij}\a _k^+,\;i,j,k\in[1;n+m]. 
& \cr
}
$$
The CAOs (13) together with (15) define completely $sl(n+1|m)$.
The outlined description via CAOs is somewhat similar to the Lie
triple system description of Lie algebras, initiated by Jacobson
[24] and generalized to Lie superalgebras by Okubo [19] (see also
[20] for further development).

The relations (15) are simple. They are however not convenient
for a quantization. It turns out that one can take only a part
of these  relations, so that they still completely define 
$sl(n+1|m)$ and are appropriate  for Hopf algebra deformations.

\smallskip\n
{\it Proposition 1} [7]. $U[sl(n+1|m)]$ is an associative unital 
superalgebra with generators 
$\a_i^{\pm}, \;\; i\in [1;n+m]$ and relations:
$$
\eqalignno{
& \[ \a_1^\xi ,\a _2^\xi \]=0,  \quad 
\[ a_1^\xi , a_1^\xi\]=0, \quad \xi=\pm, &  \cr
& \[ \[ \a_i^+ ,\a _j^- \], \a _k^+\]=\delta_{jk}\a _i^++
(-1)^{\t_i}\delta_{ij}\a _k^+,\quad & \cr
& \[ \[ \a_i^+ ,\a _j^- \], \a _k^-\]=-(-1)^{\t_{ij}\t_k}\delta_{ik}
\a _j^- -
(-1)^{\t_i}\delta_{ij}\a _k^-,\; 
   & (16) \cr
& |i-j|\leq 1, \;\;  i,j,k\in[1;n+m]. &  \cr
}
$$
The $\Z_2-$grading in $U[sl(n+1|m)]$ is induced from
$$
\deg(\a_i^\pm)=\t_i. \quad \eqno(17)
$$

Passing to the quantum case, we skip the description of
$U_q[sl(n+1|m)]$ via Chevalley generators. We write
down directly the analogue of the relations (16). To this
end introduce first the following Cartan generators:
$$
 H_i=h_1+(-1)^{\t_1}h_2+(-1)^{\t_2}h_3+\l+(-1)^{\t_{i-1}}h_i. 
 \eqno(18)
$$

\bigskip\n
{\it Theorem} [7].  $U_q[sl(n+1|m)]$ is a unital associative
algebra with  generators \hfill\break
$\{H_i,\; a_i^\pm\}_{i\in [1;n+m]}$
and relations
$$
\eqalignno{
& [H_i,H_j]=0, & (19a) \cr
& [H_i,a_j^{\pm}]=\mp(1+(-1)^{\t_i}\delta_{ij})a_j^{\pm}, 
&(19b)\cr
& \[ a_i^-, a_i^+\]={L_i-\bar{L}_i\over{q-\bar{q}}},\quad 
L_i=q^{H_i},\;\bar{L}_i\equiv L_i^{-1}=q^{-H_i},\;
\bar{q}\equiv q^{-1},  & (19c)\cr
& \[\[a_i^{\eta}, a_{i+\xi}^{-\eta}\], a_k^{\eta} 
\]_{q^{\xi (1+(-1)^{\t_i}\delta_{ik})}}=
\eta^{\t_k}\delta_{k,i+\xi}L_k^{-\xi\eta}
a_i^{\eta},  & (19d)\cr
& \[ a_1^\xi , a_2^\xi \]_q =0, \quad \[ a_1^\xi , a_1^\xi \] =0,
\quad \xi, \eta =\pm \;\;\hbox{or}\;\; \pm 1. & (19e) \cr
}
$$

\bigskip\bigskip\n
{\bf 3. Fock representations}

\bigskip\n

We proceed to describe the Fock representations of $sl(n+1|m)$ and
$U_q[sl(n+1|m)]$. The derivation, which is lengthy and nontrivial
(especially in the quantum case), will be given elsewhere.

The irreducible Fock representations are
labelled by one non-negative integer $p=1,2,\ldots$, 
called an order of the statistics. To construct them assume that the
corresponding representation space $W_p$ contains (up to
a multiple) a cyclic vector $|0\ra$, such that
$$
\eqalignno{
& a_i^-|0\ra =0,\quad i=1,2,\ldots ,n+m;&\cr
& \[ a_{ i}^-,a_{ j}^+\]|0\ra=0,\quad
i\neq j=1,2, \ldots ,n+m; & (20) \cr
& H_i|0\ra =p|0\ra, \quad i=1,2,
\ldots ,n+m.&\cr
} 
$$
Note that the above relations determine one-dimensional 
representations
(labelled by $p$) of a subalgebra, which in the nondeformed case
reduces to ${\cal B}$, Eq. (2).

{}From (19) one derives that  the deformed creation (resp. 
annihilation)
operators \hfill\break 
$q$-supercommute
$$
 \[ a_i^\xi , a_j^\xi \]_{q} =0,\quad 
\; i<j\in [1;n+m], \;\; \xi=\pm.\eqno(21)
$$
This makes evident the basis (or at least one possible basis) in a 
given Fock space, since any product of only creation operators 
can always be ordered. 

As a basis in the Fock space $W_p$ take the vectors
$$
\eqalignno{
& |p;r_1,r_2,\ldots,r_{n+m})=
\sqrt{[p-\sum_{l=1}^{n+m} r_l]!
\over {[p]![r_1]!\ldots[r_{n+m}]!}}
(a_1^+)^{r_1}(a_2^+)^{r_2}\ldots (a_n^+)^{r_n} \times  & \cr
&&\cr
& (a_{n+1}^+)^{r_{n+1}}
(a_{n+2}^+)^{r_{n+2}}
\ldots (a_{n+m}^+)^{r_{n+m}}|0\ra, \quad 
[x]={q^x-q^{-x}\over{q-q^{-1}}} & (22)\cr
}
$$
\n
with
$$
r_i\in \Z_+,\;\; i\in [1;n]; \quad r_{i}\in \{0,1\}, \;\;
i\in [n+1;n+m], \quad 
\sum_{i=1}^{n+m}r_i\leq p. \eqno(23)
$$
In order to write down the transformations of the basis under the
action of the CAOs one has to determine 
the quantum analogue of the classical 
triple relations (15). 
This actually means that one has to write down the supercommutation
relations between all Cartan-Weyl generators, expressed via
the CAOs. The latter is a necessary condition for the application of
the Poincar\'e-Birkhoff-Witt theorem, when computing the action of
the generators on the Fock basis vectors. 

Here is the result:
$$
\eqalignno{
&  L_i\L_i=\L_iL_i=1, \quad  L_iL_j=L_jL_i, \quad
L_ia_j^\pm=q^{\mp (1+ (-1)^{\t _i}\delta_{ij})}a_j^\pm L_i, & 
(24)\cr
& \[a_i^-,a_i^+\]={L_i-{\bar L}_i\over q-{\bar q}},\quad
\[a_i^\eta,a_j^\eta\]_q=0,\quad \eta=\pm,\quad i<j, & (25)  \cr 
}
$$
$$
\eqalignno{
& \[\[a_i^\eta,a_j^{-\eta}\],a_k^\eta\]_{q^{\xi(1+(-
1)^{\t_i}\d_{ik}})}=\eta^{\t_j}\d_{jk}L_k^{-\xi\eta}a_i^\eta + 
(-1)^{\t_k}\epsilon(j,k,i)
(q-\q)\[a_k^\eta,a_j^{-\eta}\]a_i^\eta&  \cr
& \quad =\eta^{\t_j}\d_{jk}L_k^{-\xi\eta}a_i^\eta + 
(-1)^{\t_k\t_j}\epsilon(j,k,i)q^\xi
(q-\q)a_i^\eta\[a_k^\eta,a_j^{-\eta}\],& (26)\cr
&\quad\quad\quad \; \xi(j-i)>0, \;
\xi,\;\eta =\pm & \cr
}
$$
where
$$
\epsilon(j,k,i)=\cases {\;\;\;1, & if $j>k>i$;\cr -1, & if 
$j<k<i$;\cr 
\;\;\; 0, & otherwise.\cr}
\eqno (27)
$$
{\it Proposition 2.} The set of all vectors (22)
constitute an orthonormal basis in  $W_p$ with respect
to the scalar product, defined in the usual way with ``bra" and
``ket" vectors and $\la 0|0 \ra =1$. 
The transformation of the basis under the
action of the CAOs reads: 
$$
\eqalignno{
& H_i|p;r_1,r_2,\ldots,r_{n+m})=\left(p-(-1)^{\t_i}r_i-
\sum_{j=1}^{n+m}r_j
\right) |p;r_1,r_2,\ldots,r_{n+m}), & (28)\cr
& a_i^-|p;r_1,\ldots,r_{n+m})=
(-1)^{\t_i(\t_1r_1+\l +\t_{i-1}r_{i-1})}q^{r_1+\ldots +r_{i-
1}}\sqrt{[r_i]
[p-\sum_{l=1}^{n+m} r_l +1]}  &\cr
& \times |p;r_1,\ldots r_{i-1},r_i-1,r_{i+1},
\ldots,r_{n+m}),& (29)   \cr
&  a_i^+|p;r_1,\ldots,r_{n+m})=
(-1)^{\t_i(\t_1r_1+\l +\t_{i-1}r_{i-1})}
\q^{r_1+\ldots +r_{i-1}}(1-\t_ir_i)
\sqrt{[r_i+1][p-\sum_{l=1}^{n+m} r_l]}&\cr
& \times |p;r_1,\ldots r_{i-1},r_i+1,r_{i+1},\ldots,r_{n+m}).& (30)   
\cr
}
$$

The Fock representations of $sl(n+1|m)$ are obtained form the
above results by replacing in Eqs. (22), (29), (30) the quantum
bracket $[..]$ with a usual bracket and setting in (29), (30)
$q=1$. The representations corresponding to an order of
statistics $p$ are irreducible and atypical representations
of the Lie superalgebra $sl(n+1|m)$. In terms of Kac's classification
[11], the Dynkin labels of the representation $W_p$ are given by
$(p,0,\ldots,0)$. This means that in general the representation $W_p$ 
is
multiply atypical [25]. More precisely, if $n\geq m$, then $W_p$ is
$m$-fold atypical; if $n<m$, then $W_p$ is $(n+1)$-fold atypical for
$p<m-n$ and $n$-fold atypical for $p\geq m-n$. 

\b
\n
{\bf 4. Properties of the underlying statistics} 
\s\n
In the present paper we have recalled
the definition of the superalgebras $sl(n+1|m)$ and
$U_q[sl(n+1|m)]$ in terms of 
creation and annihilation operators. 
Below we justify this terminology, 
illustrating on a simple example 
how each $sl(n+1|m)$ module 
$W_p$ can be viewed as a state
space, where $a_i^+$ (resp.  $a_i^-$) is interpreted as an
operator creating (resp.  annihilating) ``a particle" with, say,
energy $\varepsilon_i$. For simplicity we assume that
$n=m$. Let
$$
b_i^\pm=a_i^\pm,~~~ 
f_i^\pm=a_{i+n}^\pm,~~~ i=1,\ldots,n. \eqno(31)
$$
Consider a ``free" Hamiltonian 
$$
H=\sum_{i=1}^{n}\varepsilon_i (H_i + H_{i+n})=
\sum_{i=1}^{n}\varepsilon_i (\[b_i^+,b_i^-\]+\[f_i^+,f_i^-\]).
\eqno(32)
$$
Then 
$$ 
[H,b_i^\pm]=\pm \varepsilon_i b_i^\pm,~~~
[H,f_i^\pm]=\pm \varepsilon_i f_i^\pm. \eqno(33) 
$$ 
This result together with (nondeformed) Eqs. (29) and (30) allows one 
to interpret $r_i, ~i=1,\ldots,n$  as the number of $b-$particles 
with 
energy $\varepsilon_i$ and similarly $r_{i+n}, ~i=1,\ldots,n$  as the 
number 
of  $f-$particles with energy $\varepsilon_i$.
Then $b_i^+$ ($f_i^+$) increases this number by
one, it ``creates'' a particle in the one-particle state (=
orbital) $i$.  Similarly, the operator $b_i^-$ ($f_i^-$) diminishes 
this number by one, it ``kills'' a particle on the $i$-th orbital. On
every orbital $i$ there cannot be more than one particle of kind $f$,
whereas such restriction does not hold for the $b-$particles.
These are, so to speak, Fermi like (resp. Bose like) properties.
There is however one essential difference. If the order of the
statistics is $p$ then no more than $p$ ``particles" can be
accommodated in the system,
$$
\sum_{i=1}^{n+m}r_i\leq p.\eqno(34)
$$
This is an immediate consequence of the transformation relation
(30).  Hence the filling of a given orbital depends on how many
particles have already been accommodated on the other orbitals,
which is neither a Bose nor a Fermi-like property. For
$\sum_{i=1}^{n+m}r_i<p$ the ``particles" behave like ordinary
bosons and fermions. The maximum number of particles to be 
accommodated however in the system cannot exceed $p$. This is the
{\it Pauli principle} for this statistics.

Let us consider some configurations for $m=n=6$. Assume $p=5$.
Denote by 
$\bu$ a $b-$particle and by $\o$ an $f-$particle, and represent
the six orbibals by six boxes.

\n
1) The state
$$
|\bu\bu\bu|\bu\o\bu|~~~~~|~~~~~|~~~~~|~~~~~| 
$$
is forbidden. It is not possible to accommodate more than $p=5$
particles.

\n
2) The state
$$
|\bu\o\bu|\o\bu|~~~~~|~~~~~|~~~~~|~~~~~| 
$$
is completely filled. It contains 5 particles; no more particles
can be ``loaded" even in the empty ``boxes" (orbitals).

\n
3) The state
$$
|\bu\o\bu|\o\o|~~~~~|~~~~~|~~~~~|~~~~~| 
$$
is forbidden, because it contains two $f-$particles in the second
box.

\n
4) Consider the state
$$
|\bu\bu\o|~\bu~|~~~~~|~~~~~|~~~~~|~~~~~|. 
$$
A new $b-$particle can be accommodated in any box, whereas
the first box is ``locked" for an $f-$particle. An $f-$particle
can however be 
accommodated in any other box.

The statistics, which we have outlined above, belongs to the class
of the so-called {\it (fractional) exclusion statistics} (ES)
[26].  This issue will be considered elsewhere in more detail.
The ES was introduced in an attempt to reformulate the concept of
fractional statistics as {\it a generalized Pauli exclusion
principle} for spaces with arbitrary dimension. The literature
on the subject is vast, but practically in all publications one
studies the thermodynamics of the ES. Here we present a
microscopic description of an exclusion statistics similar as in
[27] (this is the only known to us paper attempting a microscopic
description of ES). Despite of the fact that exclusion statistics
is defined for any space dimension, so far it was applied and
tested only within 1D and 2D models. The statistics we have
studied above are examples of a microscopic description of
exclusion statistics valid in principle for any space
dimension.

\vskip 30pt
\noindent
{\bf Acknowledgments}

\s\n
N.I.S. is grateful to Prof. Randjbar-Daemi for the kind
invitation to visit the High Energy Section of the Abdus Salam
International Centre for Theoretical Physics.  T.D.P. would like
to thank the University of Ghent for a Visiting Grant, and the
Department of Applied Mathematics and Computer Science for its
kind hospitality during his stay in Ghent. The authors would also
like to thank the referee for drawing their attention to related
work.

\vskip 30pt

\n {\bf References}

\bigskip\n

{\settabs\+[11] & I. Patera, T. D. Palev, Theoretical interpretation 
of the 
   experiments on the elastic \cr 

\+ [1] & Drinfeld V G 1985 {\it DAN SSSR} {\bf 283} 1060;
         1985 {\it Sov. Math. Dokl.} {\bf 32} 254; \cr
\+     & Jimbo M 1985 {\it Lett. Math. Phys.}{\bf 10} 63 \cr 

\+ [2] & Kulish P P 1988 {\it Zapiski nauch. semin. LOMI}
         {\bf 169} 95;\cr
\+     & Kulish P P and  Reshetikhin N Yu 1989 {\it Lett. Math. 
         Phys.} 
         {\bf 18} 143; \cr
\+     & Chaichian M and Kulish P P 1990 {\it Phys. Lett.} {\bf B 
         234} 72;\cr
\+     &  Bracken A J, Gould M D and Zhang R B 1990 
         {\it Mod. Phys. Lett.} {\bf A 5} 831; \cr
\+     & Tolstoy V N 1990 {\it Lect. Notes in Physics} {\bf 370},
         Berlin, Heidelberg, New York: \cr
\+     &  Springer p. 118 \cr
\+ [3] & Palev T D 1993 {\it J. Phys. A: Math. Gen.}
         {\bf 26} L1111 and hep-th/9306016;\cr
\+     & Hadjiivanov L K 1993 {\it J. Math. Phys.} {\bf 34} 5476;\cr
\+     & Palev T D and Van der Jeugt J 1995 {\it J. Phys. A: Math. 
         Gen.} {\bf 28} 2605  and q-alg/9501020 \cr
\+ [4] & Palev T D 1994 {\it Lett. Math. Phys.} {\bf 31} 
          151  and het-th/9311163 \cr
\+ [5] & Palev T D 1998 {\it Commun. Math. Phys.} {\bf 196} 429
          and q-alg/9709003 \cr
\+ [6] & Palev T D and Parashar P 1998 {\it Lett. Math. Phys.}
          {\bf 43} 7 and q-alg/9608024   \cr
\+ [7] & Palev T D  and Stoilova N I 1999 {\it J. Math. Phys.} 
         {\bf 32} 1053 \cr

\+ [8] & Palev T D 1980 {\it J. Math. Phys.} {\bf 21} 1293 \cr

\+ [9] & Green H S 1953 {\it Phys. Rev.} {\bf 90} 270\cr 
\+ [10] & Palev T D 1982 {\it J. Math. Phys.} {\bf 23} 1100 \cr

\+ [11] & Kac V G 1979 {\it Lecture Notes in Math.} {\bf 676},
           Berlin, Heidelberg, New York:\cr
\+      &  Springer p. 597 \cr
\+ [12] & Palev T D 1979 {\it Czech. Journ. Phys.} {\bf B 29} 91\cr
\+ [13] & Palev T D 1976 Lie algebraical aspects of the quantum
          statistics {\it Thesis}  \cr
\+      & Institute for Nuclear Research and Nuclear Energy, 
          Sofia \cr          
\+ [14] & Palev T D 1977 Lie algebraical aspects of the quantum
          statistics. Unitary \cr
\+      & quantization (A-quantization) 
          {\it Preprint JINR E17-10550} and hep-th/9705032;\cr
\+      & 1980 {\it Rep. Math. Phys.} {\bf 18} 117 and 129   \cr
\+ [15] & Palev T D 1978 A-superquantization {\it Communications JINR}
          E2-11942\cr
\+ [16] & Palev T D and Van der Jeugt J {\it Fock
          representations of the Lie superalgebra} \cr
\+      & {\it  q(n+1)}, math.QA/9911176\cr
\+ [17] & Woronowicz S L 1987 {\it Publ. RIMS Kyoto University} {\bf 
23}
         117; 1987 {\it Commun. Math.}\cr
\+      & {\it Phys.} {\bf 111} 613 \cr
\+      & Pusz W and  Woronowicz S L 1989 {\it Rep. Math. Phys.} {\bf 
27} 
        231; 1989 {\it  Rep. Math. Phys.} \cr
\+      & {\bf 27} 349 \cr
\+      & Macfarlane A J 1989 {\it J. Phys. A: Math. Gen.} {\bf 22} 
4581;\cr
\+      & Biedenharn L C 1989 {\it J. Phys. } {\it A: Math. Gen.} 
{\bf 22} 
          L873; \cr
\+      & Sun C P and Fu H C 1989 {\it J. Phys.  A: Math. Gen.} 
          {\bf 22} L983 \cr
\+ [18] & Liguori A and Mintchev M 1995 {\it Commun. Math. Phys.}
      {\bf 169} 635 and hep-th/9403039; \cr
\+     &  1995 {\it Lett. Math. Phys.}  {\bf 33} 283 \cr
\+     & Loguori A, Mintchev M and Rossi M  1995 {\it Lett. Math. 
Phys.}
          {\bf 35} 163 \cr
\+  &   Zhao Liu  hep-th/9604024 \cr
\+ [19] & Okubo S 1994 {\it J. Math. Phys.} {\bf 35} 2785 \cr
\+ [20] & Meljanac S, Milekovic M and Stojic M 1998 {\it Mod.
          Phys. Lett.} {\bf A 13} 995 and \cr
\+      &  q-alg/9712017 \cr 
\+ [21] & Meljanac S, Milekovic M and  Stojic M  1999 
         {\it J. Phys.  A: Math. Gen.}
          {\bf 32} 1115 and \cr
\+      &  math-ph/9812003 \cr
\+ [22] &  Daoud M and Hassouni Y 1998 {\it  Helv. Phys. Acta} 
           {\bf 71} 599 \cr
\+ [23] & Mishra A K and Rajasekaran G 1992 {\it Pramana - J. Phys.}
          {\bf 38} L411; 1995 {\it Pramana -} \cr
\+      &  {\it J. Phys.} {\bf 45} and  hep-th/9605204 \cr
\+ [24] & Jacobson N  1949 {\it Amer. J. Math.} {\bf 71} 149 \cr

\+ [25] & Van der Jeugt J, Hughes J W B, King R C and Thierry-Mieg J  
          1990 {\it J. Math. Phys.}\cr
\+      &  {\bf 31} 2278 \cr
\+ [26] & Haldane F D M 1991 {\it Phys. Rev. Lett.} {\bf 67} 937\cr
\+ [27] & Karabali D and Nair V P 1995 {\it Nucl. Phys.}
          {\bf B 438} 551\cr

\end